\documentstyle[12pt,aps]{revtex}
\begin{document}

\title{
\begin{flushright}
\vspace{-1cm}
{\normalsize MC/TH 95/8}
\vspace{1cm}
\end{flushright}
Interpolating fields for spin-1 mesons}
\author{Michael C. Birse}
\address{Theoretical Physics Group, Department of Physics and Astronomy,\\
University of Manchester, Manchester, M13 9PL, U.K.\\}
\maketitle
\vskip 20pt
\begin{abstract}
Three commonly used types of effective theories for vector mesons are shown
to correspond to different choices of interpolating field for spin-1 particles
and the rules for transforming between them are described. The choice of fields
that transform homogeneously under the nonlinear realisation of chiral
symmetry imposes no preconceptions about the types of coupling for the mesons,
and so this representation is particularly useful for comparing different
theories. By converting them into this form, hidden-gauge theories are shown to
contain an adjustable parameter, the gauge coupling. The normal choice for this
is seen to be one that reduces the momentum dependence of the effective
$\rho\pi\pi$ coupling, explaining the success of the ``low-energy theorem" of
that approach.
\end{abstract}

\section{Introduction}

At very low energies strong interactions among pions can be described by an
effective Lagrangian based on a chirally symmetric sigma model\cite{dgh}. To
extend such a description to higher energies heavier mesons need to be
incorporated, most notably vector mesons. Various schemes for doing so have
been proposed, differing in the transformation properties of their vector
fields under chiral symmetry.

Many of these approaches are motivated by the phenomenologically successful
ideas of vector-meson dominance and universal coupling\cite{vmd}. These lead to
kinetic terms and couplings for the spin-1 mesons that have the same forms as
in a gauge theory, reflecting the assumed universal coupling of these mesons to
conserved currents. Examples include the ``massive
Yang-Mills"\cite{gg,syracuse,meissner} and ``hidden-gauge" theories\cite{bky}.
However it is not necessary to impose such a structure on the effective
Lagrangian from the start. An alternative scheme for incorporating these mesons
was suggested by Weinberg\cite{wein} and developed further by Callan, Coleman,
Wess and Zumino\cite{ccwz}. In this treatment, denoted here by WCCWZ, the
fields transform homogeneously under a  nonlinear realisation of chiral
symmetry. Another, related approach is that of Ecker {\it et al.}\ in which the
spin-1 mesons are represented by antisymmetric tensor
fields\cite{ecker1,ecker2}.

Despite the rather different forms of their Lagrangians, and the different
types of coupling contained in them, all of these approaches are in principle
equivalent. Each corresponds to a different choice of interpolating fields for
the spin-1 mesons. This is illustrated rather well in extended
Nambu--Jona-Lasinio models\cite{enjl,bijnens}, where there is considerable
freedom in the choice of auxiliary fields in the vector and axial channels. To
some extent the choice of scheme must be based on the simplicity of the
resulting Lagrangian.  In making comparisons between the approaches it is
important not to confuse features that arise from the choice of interpolating
field with those that arise from requiring, for instance, universal coupling of
the vector mesons. The former are not physical, controlling merely the
off-shell behaviour of scattering amplitudes. The latter do have physical
consequences, such as relations between on-shell amplitudes for different
processes.

Interest in these Lagrangians has recently been reawakened by the possibility
that experiments at high-luminosity accelerators such as CEBAF or DA$\Phi$NE
may be able to explore some of the couplings that have up to now been
inaccessible. Measurements of these may be able to discriminate between the
proposed effective theories\cite{bgp,esk}. These theories are also currently
being used to predict the behaviour of vector mesons in hot and dense
matter\cite{gklsy,br} and some of them lead to quite different predictions for
the mass of the $\rho$ in matter\cite{pisarski}. In both of these contexts it
is important to be able to compare theories, which may be expressed in
different formalisms, in a way which is independent of the different choices of
fields. To this end I explore here the connections between the WCCWZ,
hidden-gauge and massive Yang-Mills approaches, and their corresponding
interpolating fields.

The WCCWZ scheme, described in Sec.\ II, is particularly useful for comparisons
between theories since it imposes no prejudices about the forms of the
couplings among the mesons. As noted by Ecker {\it et al.}\cite{ecker2},
the consequences of physical assumptions like vector dominance can then be
rather transparently expressed as relations between the couplings in a WCCWZ
Lagrangian. By converting commonly used hidden-gauge and massive Yang-Mills
theories into their WCCWZ equivalents, their couplings can be directly
compared.

In the hidden-gauge approach an artificial local symmetry is introduced into
the nonlinear sigma model by the choice of field variables. The $\rho$ meson is
then introduced as a gauge boson for this symmetry. As stressed by
Georgi\cite{georgi}, the additional local symmetry has no physics associated
with it, and it can be removed by fixing the gauge. In the unitary gauge the
symmetry reduces to a nonlinear realisation of chiral symmetry, under which the
vector fields transform inhomogeneously, in contrast to those of WCCWZ.
However, with a further change of variable any vector-meson Lagrangian of the
hidden-gauge form can be converted into an equivalent WCCWZ one\cite{georgi}.
The rules for transforming a Lagrangian from hidden-gauge to WCCWZ form have
also been noted by Ecker {\it et al.}\cite{ecker2}. In Sec.\ III, this
equivalence is shown to hold for general hidden-gauge theories with axial as
well as vector mesons.

As I show here, by changing variables from the hidden-gauge to WCCWZ scheme,
the gauge coupling constant of the former is really a parameter in the choice
of interpolating vector field. This coupling does not appear in the equivalent
WCCWZ Lagrangian and so different hidden-gauge theories with different gauge
couplings can be equivalent. The conventional choice is shown to be one that
eliminates any ${\cal O}(p^3)$ $\rho\pi\pi$ coupling from the hidden-gauge
Lagrangian, so that the leading corrections to the ${\cal O} (p)$ coupling are
of order ${\cal O}(p^5)$. This reduction of the momentum dependence of the
coupling can explain why the ``low-energy  theorem" of the hidden-gauge
approach\cite{bando2,bando3} is well satisfied by the $\rho\pi\pi$ coupling
determined from the decay of on-shell $\rho$ mesons.

In an appendix I describe the relation between the WCCWZ and massive
Yang-Mills theories\cite{gg,meissner}. In the latter, the vector and axial
fields transform under a linear realisation of chiral symmetry. Three- and
four-point couplings among these fields are included and, together with the
kinetic terms, form a Yang-Mills Lagrangian with a local chiral symmetry. The
full theory does not possess this gauge symmetry since it includes mass terms
which have only global symmetry. By changing variables to spin-1 fields that
transform under the nonlinear realisation of chiral symmetry, any massive
Yang-Mills theory can be converted into an equivalent WCCWZ one.

Assumptions about the couplings implicit in commonly used hidden-gauge and
massive Yang-Mills theories can be made manifest by converting both to WCCWZ
form. The close relationship between these theories can also be seen when they
are expressed in this form. In particular both give rise to four-point
couplings that arise from assuming resonance saturation in the corresponding
scattering processes\cite{ecker1,ecker2}. In the simplest hidden-gauge
Lagrangian for pions and $\rho$ mesons\cite{bando1}, the $\rho\pi\pi$ and
$3\rho$ couplings are shown to be related by the assumed universal coupling in
these models.

\section{WCCWZ}

The WCCWZ scheme is based on the nonlinear realisation of chiral symmetry
introduced by Weinberg\cite{wein}. The starting point for this is a
two-flavour nonlinear sigma model, defined in terms of the unitary matrix
constructed out of the pion fields, $U(x)=\exp(i\hbox{\boldmath$\tau$}\cdot
\hbox{\boldmath$\pi$}(x)/f_\pi)$.
Under the global SU(2)$\times$SU(2) chiral symmetry this transforms as
$$U(x)\to g_L U(x) g_R^\dagger, \eqno(2.1)$$
with $g_L,g_R \in SU(2)$.
The nonlinear realisation of the symmetry is obtained from the
transformation properties of the square root of $U$, denoted by $u$:
$$u(x) \to g_L u(x) h^\dagger\left(u(x),g_L,g_R\right)=h\left(u(x),g_L,g_R
\right)u(x) g_R^\dagger, \eqno(2.2)$$
where $h\left(u(x),g_L,g_R\right)$ is a compensating SU(2) rotation which
depends on the pion fields at $x$ as well as $g_{L,R}$. The detailed form
of $h$ is not needed here; it can be found in Ref.\cite{ccwz}. It is
convenient to introduce the following field gradients
$$u_\mu=i (u^\dagger\partial_\mu u-u\partial_\mu u^\dagger)$$
$$\Gamma_\mu={1\over 2}(u^\dagger\partial_\mu u+u\partial_\mu u^\dagger),
\eqno(2.3)$$
which behave as an axial vector and vector respectively. Under chiral
rotations these transform as
$$u_\mu \to h u_\mu h^\dagger$$
$$\Gamma_\mu \to h\Gamma_\mu h^\dagger+h\partial_\mu h^\dagger.\eqno(2.4)$$
The quantity $u_\mu$ is seen to transform homogeneously whereas the
transformation of $\Gamma_\mu$ is inhomogeneous. In fact $\Gamma_\mu$ is the
connection on the coset space SU(2)$\times$SU(2)/SU(2) and it can be used to
construct the covariant derivative on this space:
$$\nabla_{\mu}=\partial_{\mu} + [\Gamma_\mu,\ ].\eqno(2.5)$$
The covariant derivatives of $u_\mu$ satisfy the useful relation
$$\nabla_\mu u_\nu-\nabla_\nu u_\mu=0. \eqno(2.6)$$
Also, the curvature tensor corresponding to $\Gamma_\mu$ can be expressed in
terms of $u_\mu$ as
$$\partial_\mu\Gamma_\nu-\partial_\nu\Gamma_\mu+[\Gamma_\mu,\Gamma_\nu]=
{1\over 4}[u_\mu,u_\nu],\eqno(2.7)$$

In the WCCWZ approach\cite{wein,ccwz,ecker2} vector and axial fields transform
homogeneously under this symmetry
$$V_\mu \to h V_\mu h^\dagger$$
$$ A_\mu \to h A_\mu h^\dagger,\eqno(2.8) $$
where $V_\mu={1\over 2}\hbox{\boldmath$\tau$}\cdot\hbox{\bf V}_\mu$ and
$A_\mu={1\over 2}\hbox{\boldmath$\tau$}\cdot\hbox{\bf A}_\mu$.
It is convenient to define covariant derivatives of these fields,
$$V_{\mu\nu}=\nabla_\mu V_\nu-\nabla_\nu V_\mu,\qquad
A_{\mu\nu}=\nabla_\mu A_\nu-\nabla_\nu A_\mu.\eqno(2.9)$$
A general chirally symmetric Lagrangian for $\pi\rho a_1$ physics consists of
all terms that can be constructed out of traces of products of $u_\mu$,
$V_\mu$, $A_\mu$, $V_{\mu\nu}$, $A_{\mu\nu}$ and their covariant derivatives,
and that are symmetric under parity. For example, up to fourth-order in
pion-field gradients and the vector fields, the Lagrangian includes the terms
$${\cal L}={f_\pi^2\over 4}\langle u_\mu u^\mu\rangle -{1\over 2}\langle
V_{\mu\nu}V^{\mu\nu}\rangle +m_V^2\langle V_\mu V^\mu\rangle -{i\over
2} g_1 \langle V_{\mu\nu} [u^\mu,u^\nu]\rangle  +{i\over 2} g_2 \langle
V_{\mu\nu} [V^\mu,V^\nu]\rangle$$
$$ +{1\over 8} g_3 \langle[u_\mu,u_\nu]^2\rangle  -{1\over 4} g_4 \langle
[u_\mu,u_\nu][V^\mu,V^\nu]\rangle  +{1\over 8} g_5 \langle
[V_\mu,V_\nu]^2\rangle +\cdots,\eqno(2.10)$$
where $\langle \cdots\rangle $ denotes a trace in SU(2) space. The
terms written out explicitly here are the ones we shall need in discussing the
connection to the hidden-gauge Lagrangian of Bando {\it et al.}\cite{bando1}.
These include the famous $\langle[u_\mu,u_\nu]^2\rangle$ term introduced by
Skyrme to stabilise solitons in a nonlinear sigma model\cite{skyrme}. Obviously
many other three- and four-point interactions, involving the axial as well as
the vector field, should also be present in the full effective Lagrangian.

As pointed out by Kalafatis\cite{kal} and discussed further
elsewhere\cite{kb}, the four-point couplings must satisfy
inequalities relating them to the three-point couplings if the Hamiltonian
corresponding to (2.10) is to be bounded from below. For example the
coefficient
of the Skyrme term should satisfy $g_3\geq g_1^2$. If one assumes that vector
dominance holds in the strong interaction and that the scattering processes
corresponding to these four-point couplings are saturated by exchange of a
single resonance, in this case the $\rho$, then the equalities hold, for
example $g_3=g_1^2$, as discussed in\cite{ecker2}.

\section{Hidden-gauge theories}

In the simplest version of the hidden-gauge approach, the gauge symmetry is
just SU(2) and only vector mesons are treated as gauge bosons\cite{bando1}.
The extension to axial-vector mesons requires a local SU(2)$\times$SU(2)
symmetry, which can be introduced by writing $U(x)$ as a product of three
unitary matrices\cite{bando2,meissner,bky},
$$U(x)=\xi_L(x)^\dagger \xi_M(x) \xi_R(x). \eqno(3.1)$$
Since at each point in space-time this factorisation is arbitrary, the new
variables are symmetric under
$$\xi_R(x)\to h_R(x)\xi_R(x)g_R^\dagger$$
$$\xi_L(x)\to h_L(x)\xi_L(x)g_L^\dagger$$
$$\xi_M(x)\to h_L(x)\xi_M(x)h_R^\dagger(x),\eqno(3.2)$$
where $h_{L,R}(x)$ are SU(2) matrices with arbitrary $x$-dependence. The
freedom to make space-time dependent rotations of $\xi_{r,l,m}(x)$ in this way
provides the local SU(2)$\times$SU(2) symmetry of this scheme.

One can always to choose to work in the unitary gauge where
$$\xi_R(x)=\xi_L^\dagger(x)=u(x),\qquad \xi_M(x)=1, \eqno(3.3)$$
for all $x$. The symmetry (3.1) then reduces to the usual nonlinear realisation
of chiral symmetry\cite{wein,ccwz}, as in Eq.~(2.2), where the $x$ dependence
of $h_R(x)=h_L(x)$ is no longer arbitrary but is given in terms of the pion
fields. This gauge fixing thus provides the basis for translating between the
hidden-gauge and WCCWZ formalisms.

In this approach, spin-1 fields are introduced as gauge bosons of this
artificial local symmetry. Right- and left-handed gauge fields transform under
the symmetry as, respectively,
$$\widehat X_\mu(x)\to h_R(x)\widehat X_\mu(x)h_R^\dagger(x)+{i\over
{\sqrt 2}g}h_R(x)\partial_\mu(x)h_R^\dagger(x)$$
$$\widehat Y_\mu(x)\to h_L(x)\widehat Y_\mu(x)h_L^\dagger(x)+{i\over
{\sqrt 2}g}h_L(x)\partial_\mu(x)h_L^\dagger(x),\eqno(3.4)$$
where I use hats to distinguish the hidden-gauge spin-1 fields from those of
the WCCWZ approach. The corresponding gauge-covariant field strengths are
$$\widehat X_{\mu\nu}= \partial_\mu \widehat X_\nu-\partial_\nu \widehat X_\mu
-i{\sqrt 2}g[\widehat X_\mu,\widehat X_\nu]$$
$$\widehat Y_{\mu\nu}= \partial_\mu \widehat Y_\nu-\partial_\nu \widehat Y_\mu
-i{\sqrt 2}g[\widehat Y_\mu,\widehat Y_\nu].\eqno(3.5)$$
It is usually more convenient to work in terms of the vector and axial fields,
$\widehat V_\mu=(\widehat X_\mu+\widehat Y_\mu)/\sqrt 2$ and $\widehat
A_\mu=(\widehat X_\mu-\widehat Y_\mu)/\sqrt 2$. The field strengths for these
are
$$\widehat V_{\mu\nu}= \partial_\mu \widehat V_\nu-\partial_\nu \widehat V_\mu
-ig[\widehat V_\mu,\widehat V_\nu]-ig[\widehat A_\mu,\widehat A_\nu]$$
$$\widehat A_{\mu\nu}= \partial_\mu \widehat A_\nu-\partial_\nu \widehat A_\mu
-ig[\widehat V_\mu,\widehat A_\nu]-ig[\widehat A_\mu,\widehat V_\nu].
\eqno(3.6)$$

The gauge-covariant first derivatives of the pion fields are
$$R_\mu= -i\left[(\partial_\mu\xi_L)\xi^\dagger_L
-i{\sqrt 2}g\widehat X_\mu\right]$$
$$L_\mu= -i\left[(\partial_\mu\xi_R)\xi^\dagger_R-i{\sqrt 2}g\widehat
Y_\mu\right]$$
$$M_\mu= -i\left[(\partial_\mu\xi_M)\xi^\dagger_M+i{\sqrt 2}g\xi_M\widehat
X_\mu\xi^\dagger_M-i{\sqrt 2}g\widehat Y_\mu\right].\eqno(3.7)$$
Of these, $R_\mu$ transforms covariantly under the right-handed local symmetry,
$L_\mu$ and $M_\mu$ under the left-handed. A general gauge-invariant Lagrangian
in this approach consists of all terms that can be constructed out of traces
of products of $R_\mu$, $L_\mu$, $M_\mu$, $\widehat X_{\mu\nu}$, $\widehat
Y_{\mu\nu}$, and their covariant derivatives, and that are symmetric under
parity. Factors of $\xi_M$ and $\xi^\dagger_M$ should be inserted between
right- and left-covariant quantities. Writing out explicitly only terms of
second order in the pion field gradients (which also provide mass terms for the
heavy mesons) and the vector-meson kinetic terms, one has the
Lagrangian\cite{bando2,meissner,bky}
$${\cal L}= {af_\pi^2\over 4}\langle (L_\mu+\xi_M R_\mu \xi^\dagger_M)^2\rangle
+{bf_\pi^2\over 4}\langle (L_\mu-\xi_M R_\mu \xi^\dagger_M)^2\rangle$$
$$+{cf_\pi^2\over 4}\langle M_\mu M^\mu\rangle +{df_\pi^2\over 4}\langle
(L_\mu-\xi_M R_\mu \xi^\dagger_M-M_\mu)^2\rangle$$
$$-{1\over 2}\langle \widehat X_{\mu\nu}\widehat X^{\mu\nu}+\widehat Y_{\mu\nu}
\widehat Y^{\mu\nu}\rangle +\cdots.\eqno(3.8)$$

In the unitary gauge defined by (3.3), the transformation
properties of the spin-1 fields are
$$\widehat V_\mu \to h \widehat V_\mu h^\dagger+{i\over g}h\partial_\mu
h^\dagger$$
$$\widehat A_\mu \to h \widehat A_\mu h^\dagger,\eqno(3.9)$$
where $h$ is the compensating SU(2) rotation of the nonlinear realisation
of chiral symmetry, Eq.~(2.2). The covariant gradients of Eq.~(3.7) reduce to
$$R_\mu=iu^\dagger\partial_\mu u-g(\widehat V_\mu+\widehat A_\mu)$$
$$L_\mu=iu\partial_\mu u^\dagger-g(\widehat V_\mu-\widehat A_\mu)$$
$$M_\mu=2g\widehat A_\mu.\eqno(3.10)$$
Since these can be combined to give
$$R_\mu+L_\mu=2(i\Gamma_\mu-g\widehat V_\mu)$$
$$R_\mu-L_\mu=u_\mu-2g\widehat A_\mu,\eqno(3.11)$$
we see that $\widehat V_\mu$ always appears in the combination
$$V_\mu=\widehat V_\mu-{i\over g}\Gamma_\mu. \eqno(3.12)$$
This transforms homogeneously under the nonlinear chiral rotation, as can be
seen from (2.4) and (3.9).

In the unitary gauge we can therefore change variables to the vector field
$V_\mu$ of Eq.~(3.12) to obtain a Lagrangian of the WCCWZ type. (The axial
field already transforms homogeneously in this gauge, Eq.~(3.9).) With the aid
of Eq.~(2.7), the field strengths of Eq.~(3.6) can be expressed in terms of the
new field as
$$\widehat V_{\mu\nu}= V_{\mu\nu}+{i\over 4g}[u_\mu,u_\nu]-ig[V_\mu,V_\nu]
-ig[A_\mu,A_\nu]$$
$$\widehat A_{\mu\nu}= A_{\mu\nu}-ig[V_\mu,A_\nu]-ig[A_\mu,V_\nu],\eqno(3.13)$$
where the covariant field gradients are defined in (2.9) above. Terms involving
higher gauge-covariant derivatives can be rewritten in terms of the covariant
derivative (2.5) using
$$D_\mu=\partial_\mu-ig[\widehat V_\mu,\ ]=\nabla_\mu-ig[V_\mu,\ ].
\eqno(3.14)$$
Each term of the general hidden-gauge Lagrangian in the unitary gauge has a
corresponding term in the general WCCWZ Lagrangian, where $\widehat V_\mu
-i\Gamma_\mu/g$ has been replaced by $V_\mu$, $D_\mu$ by $\nabla_\mu$,
$\widehat V_{\mu\nu}$ by $V_{\mu\nu}$, and $\widehat A_{\mu\nu}$ by
$A_{\mu\nu}$. The coupling constants will not be identical but, if one takes
account of Eqs.~(3.13, 14), there is a well defined way to convert from one
approach to the other. This generalises Georgi's observation\cite{georgi} of
the equivalence of the two formalisms to the case of axial as well as vector
fields.

An important feature to note is that the gauge coupling constant $g$ of the
hidden-gauge approach does not appear in the WCCWZ approach. Indeed different
hidden-gauge Lagrangians with different values of $g$ can be equivalent to the
same WCCWZ theory. This should not be too surprising: the local symmetry is not
physical but arises from a particular choice of field variables in Eq.~(3.1)
and hence the corresponding coupling is not a physical quantity. The
significance of $g$ becomes clearer if one starts from a WCCWZ Lagrangian and
converts it into a hidden-gauge one using Eq.~(3.12) in reverse. Any value of
$g$ can be used in (3.12) to define a new vector field $\widehat V_\mu$ and the
resulting Lagrangian will have the form of a hidden-gauge theory in the unitary
gauge. Different choices of $g$ therefore correspond to different choices of
interpolating vector field. The value of $g$ should thus be fixed by
considerations of calculational convenience, for example the elimination of
certain types of term from the effective Lagrangian.

To explore this equivalence in more detail, let us examine a specific
hidden-gauge theory. The example considered is the most commonly used
hidden-gauge model, introduced by Bando {\it et al.}\cite{bando1}. This
contains a vector but no axial field and so is invariant under the diagonal
SU(2) subgroup of the local symmetry only. Its Lagrangian has the form
$${\cal L}={f_\pi^2\over 4}\langle (L_\mu-R_\mu)^2\rangle
+{af_\pi^2\over 4}\langle (L_\mu+R_\mu)^2\rangle
-{1\over 2}\langle \widehat V_{\mu\nu}\widehat V^{\mu\nu}\rangle ,\eqno(3.15)$$
where
$$R_\mu= -i\left[(\partial_\mu\xi_L)\xi^\dagger_L
-ig\widehat V_\mu\right]$$
$$L_\mu= -i\left[(\partial_\mu\xi_R)\xi^\dagger_R-ig\widehat V_\mu\right].
\eqno(3.16)$$
In the unitary gauge this becomes
$${\cal L}={f_\pi^2\over 4}\langle u_\mu u^\mu\rangle
+af_\pi^2\langle (i\Gamma_\mu-g\widehat V_\mu)^2\rangle
-{1\over 2}\langle \widehat V_{\mu\nu}\widehat V^{\mu\nu}\rangle, \eqno(3.17)$$
showing that the $\rho$ mass is given in terms of the parameter $a$ by
$$m_V^2=ag^2f_\pi^2. \eqno(3.18)$$
Using (3.12, 13) this can be expressed in the form of the WCCWZ Lagrangian
of Eq.~(2.10), with the following values for the coupling constants:
$$g_1={1\over 2g},\quad g_2=2g,\quad g_3={1\over 4g^2},\quad g_4=1,\quad
g_5=4g^2.
\eqno(3.19)$$
The couplings in this model thus satisfy the relations
$$g_3=g_1^2,\quad g_4=g_1g_2,\quad g_5=g_2^2, \eqno(3.20)$$
which arise from assuming vector meson dominance in the strong interaction and
saturation of the four-point couplings by the $\rho$-meson alone, as discussed
following Eq.~(2.10). The other condition that defines the model is a relation
between the $\rho\pi\pi$ and $3\rho$ couplings,
$$g_1={1\over g_2}. \eqno(3.21)$$
These relations (3.20, 21) allow the three- and four-point couplings to be
combined into a kinetic term for the vector field with a Yang-Mills form.
Note that all of these relations hold for any value of the $\rho$ mass (or
equivalently of the parameter $a$).

Complete vector dominance of the pion electromagnetic coupling is obtained in
this model if the $\rho$ mass satisfies the KSRF relation\cite{ksrf} (in its
second form)
$$m_V^2=2g^2f_\pi^2, \eqno(3.22)$$
In terms of the parameters of Bando {\it et al.}\cite{bando1,bky} this
corresponds to $a=2$. Vector dominance in the couplings of the photon to all
hadrons requires universal coupling of the $\rho$ meson to the conserved
isospin current. By examining the $\rho\pi\pi$ coupling contained in (3.17),
$$-2igaf_\pi^2\langle \widehat V^\mu \Gamma_\mu\rangle ={1\over 2}ag
\widehat{\hbox{\bf V}}^\mu\cdot\hbox{\boldmath$\pi$}\wedge\partial_\mu
\hbox{\boldmath$\pi$}+{\cal O}(\pi^4), \eqno(3.23)$$
and the $3\rho$ coupling
$$2ig\langle (\partial_\mu\widehat V_\nu-\partial_\nu\widehat V_\mu)
[\widehat V^\mu,\widehat V^\nu]\rangle =-g\widehat{\hbox{\bf V}}^\mu\cdot
\widehat{\hbox{\bf V}}^\nu\wedge\partial_\mu \widehat{\hbox{\bf V}}_\nu.
\eqno(3.24)$$
we see that for $a=2$ the couplings have the same strength and so the model
embodies universal coupling of the $\rho$ to itself and to the  pion.

In looking at predictions of the model (3.15), consequences of the hidden-gauge
choice of interpolating field should not be confused with those arising from
the relations between the coupling constants (3.20, 21). The former controls
merely the form of off-shell extrapolations of those amplitudes. The latter
lead to relations between amplitudes for physical processes, and are of
course specific to the choice of Lagrangian. For example the relation (3.21)
between the $\rho\pi\pi$ and $3\rho$ couplings can be removed without violating
the hidden-gauge invariance by adding a term of the form $\langle \widehat
V_{\mu\nu} \left[R^\mu+L^\mu,R^\nu+L^\nu\right]\rangle$ to the Lagrangian
(3.15). This is invariant under the same local SU(2) symmetry as the rest of
the Lagrangian. It provides an additional contribution to the $3\rho$ coupling
beyond that in the kinetic term. After gauge-fixing and change of variables,
such a term would lead to an equivalent WCCWZ Lagrangian that would not satisfy
(3.21).

To see why the specific choice of field in (3.15) is particularly convenient,
consider the $\rho\pi\pi$ of the corresponding WCCWZ Lagrangian (2.10),
$$-{i\over 2} g_1 \langle V_{\mu\nu} [u^\mu,u^\nu]\rangle =g_1\partial^\mu
\hbox{\bf V}^\nu\cdot\partial_\mu\hbox{\boldmath$\pi$}\wedge\partial_\nu
\hbox{\boldmath$\pi$}, \eqno(3.25)$$
This is of third-order in the momenta. In contrast, when the model is expressed
in hidden-gauge form (3.15), the leading $\rho\pi\pi$ coupling (3.23) is of
first order in the momenta of the particles involved. Using the interpolating
field defined by (3.12) with the constant $g$ given by
$$g={1\over 2g_1} \eqno(3.26)$$
eliminates any ${\cal O}(p^3)$ term from the hidden-gauge Lagrangian. This
happens even if the initial WCCWZ Lagrangian contains other, higher-derivative
$\rho\pi\pi$ couplings. The advantage of the hidden-gauge choice of
interpolating field with this $g$ is that any corrections to the leading
$\rho\pi\pi$ of Eq.~(3.23) are at least of order ${\cal O}(p^5)$. Provided
$m_V$ is small compared with the scale at which physics beyond the $\pi\rho$
Lagrangian becomes significant, the momentum dependence of the effective
$\rho\pi\pi$ coupling should be small in the hidden-gauge representation. This
explains why the KSRF relation\cite{ksrf} in its first version, which relates
the $\rho\pi\pi$ coupling and $\gamma\rho$ mixing at zero four-momentum and
forms the ``low-energy theorem" of the hidden-gauge
approach\cite{bando2,bando3}, is actually rather well satisfied by the values
for on-shell $\rho$ mesons. The freedom to choose $g$ in this way after
renormalisation can also explain why the low-energy theorem continues to hold
when loop corrections are included\cite{hy,hky}.

\section{Conclusions}

As described here any effective theories of spin-1 mesons and pions can be
expressed in either WCCWZ, hidden-gauge or massive Yang-Mills form. These
formalisms correspond to different choices of interpolating fields for the
spin-1 mesons. The rules for transforming a theory from one form to another
generalise the equivalences between the approaches that have previously been
noted for particular cases. Since all the formalisms are equivalent, the choice
between them must depend on the convenience of the corresponding Lagrangians
for a specific calculation.

The hidden-gauge case is of particular interest since it involves a vector
field that  depends on a continuous parameter $g$, which acts as the gauge
coupling for the local symmetry introduced in this approach. This parameter can
be chosen to remove the ${\cal O}(p^3)$ momentum dependence of the $\rho\pi\pi$
coupling, and indeed this choice is implicitly used in applications of the
hidden-gauge approach. This reduction of the momentum dependence of the
$\rho\pi\pi$ coupling can explain the success of the ``low-energy theorem"
of this approach. It suggests that this may be a particularly convenient
representation to use for low-energy $\pi\rho$ physics.

The massive Yang-Mills representation (discussed in the Appendix) leads to no
such simplifications. It is however based on spin-1 fields that transform
linearly under chiral rotations. This may be of use in studying the
restoration of chiral symmetry if Weinberg's ideas of ``mended" chiral
symmetry\cite{mend} are relevant in this context, as has been suggested by
Brown and Rho\cite{br}.

The WCCWZ formalism is particularly useful as framework for comparing different
theories and for elucidating the implicit relations between their couplings. It
shows that commonly used hidden-gauge and massive Yang-Mills theories both give
rise to four-point couplings that arise from assuming resonance saturation in
the corresponding scattering processes\cite{ecker1,ecker2}. In the hidden gauge
case the $\rho\pi\pi$ and $3\rho$ couplings are related by the assumption of
universal coupling. Although this is not the case in the simplest massive
Yang-Mills theory, non-minimal terms can be added to such a theory so that it
leads to similar predictions. Such assumptions about the couplings in these
models could be tested experimentally using processes that are sensitive to
the $3\rho$ coupling, for example $\rho\rightarrow\pi^+\pi^-2\pi^0$ \cite{esk}.

\section*{Acknowledgments}

I am grateful to D. Kalafatis for extensive discussions at the start of this
work. I would like to thank R. Plant for useful discussions and J. McGovern
for critically reading the manuscript. I am also grateful to the Institute for
Nuclear Theory, University of Washington, Seattle for its hospitality while
part of this work was completed. This work is supported by the EPSRC and PPARC.

\section*{Appendix: Massive Yang-Mills}

The massive Yang-Mills approach\cite{gg,syracuse,meissner} is based on vector
and axial fields that transform linearly under the SU(2)$\times$SU(2)
symmetry. The right- and left-handed combinations of these spin-1 fields,
denoted here by
$\widetilde X_\mu$ and $\widetilde Y_\mu$, transform as
$$\widetilde X_\mu\rightarrow g_R\widetilde X_\mu g_R^\dagger$$
$$\widetilde Y_\mu\rightarrow g_L\widetilde Y_\mu g_L^\dagger. \eqno(A.1)$$
The Lagrangian for these is chosen to contain kinetic terms of the Yang-Mills
form, including three- and four-point interactions. The couplings of the spin-1
fields to pions are also chosen to have a gauge-invariant form, ensuring
universal coupling of the $\rho$ and allowing photons to be coupled in a way
consistent with vector dominance. Although the interaction terms respect a
local SU(2)$\times$SU(2) symmetry, the full theory does not since it also
includes mass terms for the spin-1 mesons.

A simple massive Yang-Mills Lagrangian, which illustrates the features of the
approach, consists of the gauged sigma model (a nonlinear version of the model
used by Gasiorowicz and Geffen\cite{gg})
$${\cal L}={f_0^2\over 4}\langle\widetilde D_\mu U(\widetilde D_\mu U)^\dagger
\rangle-{1\over 2}\langle\widetilde X_{\mu\nu}\widetilde X^{\mu\nu}
+\widetilde Y_{\mu\nu}\widetilde Y^{\mu\nu}\rangle$$
$$+m_V^2\langle \widetilde X_\mu\widetilde X^\mu
+\widetilde Y_\mu\widetilde Y^\mu\rangle, \eqno(A.2)$$
where
$$\widetilde D_\mu U=\partial_\mu U+i\sqrt 2\widetilde gU\widetilde X_\mu
-i\sqrt 2\widetilde g\widetilde Y_\mu U, \eqno(A.3)$$
and the field strengths $\widetilde X_{\mu\nu}$, $\widetilde Y_{\mu\nu}$
are defined analogously to those in Eq.~(3.5).

The massive Yang-Mills theory can be converted into an equivalent WCCWZ one by
using $u(x)$ to construct spin-1 fields that transform under the nonlinear
realisation of chiral symmetry (2.8):
$$X_\mu=u\widetilde X_\mu u^\dagger$$
$$Y_\mu=u^\dagger\widetilde Y_\mu u. \eqno(A.4)$$
The kinetic terms can be expressed in terms of the covariant derivatives of
these fields, defined as in (2.9), using
$$\widetilde X_{\mu\nu}=u^\dagger\Biggl[X_{\mu\nu}+{i\over 2}[u_\mu,X_\nu]
-{i\over 2}[u_\nu,X_\mu]-i\sqrt 2\widetilde g[X_\mu,X_\nu]\Biggr]u$$
$$\widetilde Y_{\mu\nu}=u^\dagger\Biggl[Y_{\mu\nu}-{i\over 2}[u_\mu,Y_\nu]
+{i\over 2}[u_\nu,Y_\mu]-i\sqrt 2\widetilde g[Y_\mu,Y_\nu]\Biggr]u.\eqno(A.5)$$
In terms of $u_\mu$ and these fields, the pion kinetic term can be written
$$\langle\widetilde D_\mu U(\widetilde D_\mu U)^\dagger\rangle=\langle[u_\mu
-\sqrt 2\widetilde g(X_\mu-Y_\mu)]^2\rangle.\eqno(A.6)$$
This contains a $\pi a_1$ mixing term which can be removed by an appropriate
shift in the definition of the axial field\cite{gg,ecker2}. It is thus
convenient to define WCCWZ vector and axial fields by
$$V_\mu={1\over \sqrt 2}(X_\mu+Y_\mu)$$
$$A_\mu={1\over \sqrt 2}(X_\mu-Y_\mu)-{\widetilde g f_0^2\over 2m_A^2}u_\mu.
\eqno(A.7)$$

The kinetic terms for the spin-1 fields can then be expressed in terms of
$V_\mu$ and $A_\mu$ and their covariant derivatives (2.9), making use of (2.6).
The Lagrangian (A.2) then takes the form
$${\cal L}={f_\pi^2\over 4}\langle u_\mu u^\mu\rangle+m_V^2\langle V_\mu
V^\mu\rangle+m_A^2\langle A_\mu A^\mu\rangle$$
$$-{1\over 2}\Bigg\langle\Bigg\{V_{\mu\nu}-i\widetilde g[V_\mu,V_\nu]
-i\widetilde g[A_\mu,A_\nu]$$
$$+i{\textstyle{1\over 2}}Z^2\Bigl([u_\mu,A_\nu]-[u_\nu,A_\mu]\Bigr)$$
$$+{i\over 4\widetilde g}(1-Z^4)[u_\mu,u_\nu]\Bigg\}^2\Bigg\rangle$$
$$-{1\over 2}\Big\langle\Big\{A_{\mu\nu}-i\widetilde g[V_\mu,A_\nu]
-i\widetilde g[A_\mu,V_\nu]$$
$$+i{\textstyle{1\over 2}}Z^2\Bigl([u_\mu,V_\nu]
-[u_\nu,V_\mu]\Bigr)\Big\}^2\Big\rangle. \eqno(A.8)$$
where
$$Z^2=1-{\widetilde g^2 f_0^2\over m_A^2}=1-{\widetilde g^2 f_\pi^2\over
m_V^2}, \eqno(A.9)$$
and the physical pion decay constant is given by\cite{meissner}
$$f_\pi^2=f_0^2Z^2, \eqno(A.10)$$
and the $a_1$ mass by
$$m_A^2=m_V^2/Z^2. \eqno(A.11)$$

Although I have demonstrated the equivalence here for only the theory defined
by (A.2), it is general. Any massive Yang-Mills Lagrangian can be expressed
in WCCWZ form using (A.4, 5, 7). Conversely, any WCCWZ Lagrangian can be
converted into an equivalent massive Yang-Mills theory by inverting these
changes of variable. Of course the resulting Lagrangian can contain many terms
beyond those present in (A.2), including many non-gauge-invariant
interactions. Combined with the results of Section III, this reproduces and
generalises the well-known equivalence of the hidden-gauge and massive
Yang-Mills formalisms\cite{schechter,mz,yamaw,meissner}.

By comparing the terms in the Lagrangian (A.8) with the corresponding ones in
(2.10), we can see that the couplings satisfy the relations (3.20) arising from
assuming resonance saturation of the four-point interactions. This is similar
to the hidden-gauge theory defined by (3.15). The two theories are thus closely
related although obviously not identical: the massive Yang-Mills one contains
an axial as well as a vector field, and its $\rho\pi\pi$ and $3\rho$ couplings
do not satisfy (3.21). The latter is a consequence of the the additional
momentum-dependent $\rho\pi\pi$ couplings that appear after diagonalising in
the $\pi a_1$ sector. One can always cancel out this momentum dependence by
adding extra ``nonminimal" terms to the massive Yang-Mills
Lagrangian\cite{syracuse}.\footnote{One can also force the theory into exact
equivalence with the one of Bando {et al.}\cite{bando1} by imposing a suitable
constraint on the axial field\cite{ks,meissner}. This can be seen using the
WCCWZ form (A.8): if one demands that $A_\mu=(Z^2/2g)u_\mu$ then one is left
with the WCCWZ equivalent of (3.16).} The resulting massive Yang-Mills
Lagrangian is then exactly equivalent to the hidden-gauge one involving
axial as well as vector mesons introduced in Ref.~\cite{bando3}.

Finally, for completeness, I should mention the approach suggested by Brihaye,
Pak and Rossi\cite{bpr} and investigated further by Kuraev, Silagadze and
coworkers\cite{kuraev,esk}. This is based on a Yang-Mills-type coupling of the
$\rho$, as in the Lagrangian (A.2), but without a chiral partner $a_1$-field.
Simply omitting the axial field from that Lagrangian leaves a theory that is
not chirally symmetric. However as described in Ref.\cite{bpr} additional
counterterms can be added to that Lagrangian to ensure that low-energy theorems
arising from chiral symmetry are maintained.

Such a theory can be generated by taking a hidden-gauge Lagrangian, such
as that of Eq.~(3.17), and reversing the procedure above for converting fields
that transform linearly under chiral symmetry into ones that transform
nonlinearly. Specifically one can define a new vector field $\widetilde V_\mu$,
related to the hidden-gauge field $\widehat V_\mu$ by
$$\widehat V_\mu={1\over 2}\left(u^\dagger \widetilde V_\mu u
+u\widetilde V_\mu u^\dagger\right). \eqno(A.12)$$
Note that this $\widetilde V_\mu $ has no axial partner and so does not
transform in any simple way under chiral transformations. By adding and
subtracting suitable terms, similar to those in Eq.~(A.6), the pion kinetic
term can be converted to a form involving gauge-covariant derivatives. The full
Lagrangian is then
$${\cal L}={f_\pi^2\over 4}\langle \widetilde D_\mu U(\widetilde D_\mu
U)^\dagger\rangle+{\widetilde gf_\pi^2\over 2}\langle\widetilde V_\mu
(uu_\mu u^\dagger-u^\dagger u_\mu u)\rangle-{\widetilde g^2f_\pi^2\over 4}
\langle (u^\dagger V_\mu u-u V_\mu u^\dagger)^2\rangle$$
$$+af_\pi^2\langle (i\Gamma_\mu-g\widehat V_\mu)^2\rangle
-{1\over 2}\langle \widehat V_{\mu\nu}\widehat V^{\mu\nu}\rangle.
\eqno(A.13)$$
By choosing the coupling $\widetilde g$ to be related to the parameters
of the hidden-gauge Lagrangian by
$$\widetilde g={ag\over 2}, \eqno(A.14)$$
one can ensure that the ${\cal O}(p)$ $\rho\pi\pi$ coupling (3.23) in the
fourth term of (A.13) is cancelled by that in the second term. The
$\rho\pi\pi$ coupling is then given entirely by the gauge-covariant
derivatives in the first term of (A.13).

The resulting theory has a Yang-Mills structure for the $\rho$ kinetic energy
and $\rho\pi\pi$ couplings, together with a $\rho$ mass term and a number of
additional couplings. These extra couplings are required if the low-energy
theorems of chiral symmetry are to be satisfied. They include the counterterms
discussed in Refs.\cite{bpr,kuraev} together with many others. For example the
third term of (A.13) contains a momentum-independent $\rho\rho\pi\pi$
coupling. This term is omitted in the calculations of Kuraev {\it et
al.}\cite{kuraev,esk} but it is needed to cancel out a corresponding piece of
the pion kinetic term, which would otherwise give a nonzero amplitude for
$\pi\rho$ scattering at threshold in the chiral limit.

\end{document}